# LSSVM-ABC Algorithm for Stock Price prediction

Osman Hegazy [1], Omar S. Soliman [2] and Mustafa Abdul Salam[3]
[1, 2](Faculty of Computers and Informatics, Cairo University, Egypt)
[3](Higher Technological Institute (H.T.I), 10th of Ramadan City, Egypt)

**ABSTRACT :** In this paper, Artificial Bee Colony (ABC) algorithm which inspired from the behavior of honey bees swarm is presented. ABC is a stochastic population-based evolutionary algorithm for problem solving. ABC algorithm, which is considered one of the most recently swarm intelligent techniques, is proposed to optimize least square support vector machine (LSSVM) to predict the daily stock prices. The proposed model is based on the study of stocks historical data, technical indicators and optimizing LSSVM with ABC algorithm. ABC selects best free parameters combination for LSSVM to avoid over-fitting and local minima problems and improve prediction accuracy. LSSVM optimized by Particle swarm optimization (PSO) algorithm, LSSVM, and ANN techniques are used for comparison with proposed model. Proposed model tested with twenty datasets representing different sectors in S&P 500 stock market. Results presented in this paper show that the proposed model has fast convergence speed, and it also achieves better accuracy than compared techniques in most cases.

## I. INTRODUCTION

The stock market field is characterized by unstructured nature, data intensity, noise, and hidden relationships. Also the behavior of a stock time series is close to random. For these reasons the stock market prediction is not a simple task. In field of stock market, there are several prediction tools that have been used since years ago. Fundamental and technical analyses were the first two methods used to forecast stock prices. ARIMA (Autoregressive Integrated Moving Average) is one of the most commonly used time series prediction methods [1]. ARIMA method deals with linear, but it is difficult to deal with nonlinear feature in time series [2]. GARCH (Generalized Autoregressive Conditional Heteroskedasticity) linear time series prediction method is also used [3]. Artificial Neural Networks (ANNs) tool which is a branch of Computational Intelligence (CI) is become one of the most commonly machine learning techniques used in stock market prediction. ANNs prediction method is used to overcome the limitations of above methods. In most cases ANNs suffer from over-fitting problem due to the large number of parameters to fix, and the little prior user knowledge about the relevance of the inputs in the analyzed problem [4]. Support vector machines (SVMs) have been developed as an alternative that avoids the above prediction models limitations. Their practical successes can be attributed to solid theoretical foundations based on VC-theory [5]. SVM computes globally optimal solutions, unlike those obtained with ANN, which tend to fall into local minima [6]. Least squares support vector machine (LSSVM) method which is presented in [7], is a reformulation of the traditional SVM algorithm. LSSVM uses a regularized least squares function with equality constraints, leading to a linear system which meets the Karush-Kuhn-Tucker (KKT) conditions for obtaining an optimal solution. Although LSSVM simplifies the SVM procedure, the regularization parameter and the kernel parameters play an important role in the regression system. Therefore, it is necessary to establish a methodology for properly selecting the LSSVM free parameters, in such a way that the regression obtained by LSSVM must be robust against noisy conditions, and it does not need priori user knowledge about the influence of the free parameters values in the problem studied [8].

The perceived advantages of evolutionary strategies as optimization methods motivated the authors to consider such stochastic methods in the context of optimizing SVM. A survey and overview of evolutionary algorithms (EAs) is found in [9]. Since 2005, D. Karaboga and his research group have been studying the Artificial Bee Colony (ABC) algorithm and its applications to real world problems. Karaboga and Basturk have investigated the performance of the ABC algorithm on unconstrained numerical optimization problems which is found in [10], [11], [12] and its extended version for the constrained optimization problems in [13]. Neural network trained with ABC algorithm is presented in [14], [15]. ABC Algorithm based approach for structural optimization is presented in [16]. Applying ABC algorithm for optimal multi-level thresholding is introduced in [17]. ABC was used for MR brain image classification in [18], cluster analysis in [19], face pose estimation in [20], and 2D protein folding in [21]. A new hybrid ABC algorithm for robust optimal design and manufacturing is introduced in [22]. Hybrid artificial bee colony-based approach for optimization of multi-pass turning operations is used in [23]. Intrusion detection in AODV-based MANETs using ABC and negative selection algorithms is presented in [24]. A study on Particle Swarm Optimization (PSO) and ABC algorithms for multilevel thresholding is introduced in [25]. Heuristic approach for inverse kinematics of robot arms is found in [26]. A combinatorial ABC Algorithm applied for traveling salesman problem is introduced in [27]. Identifying nuclear power plant transients using the discrete binary ABC Algorithm is presented in [28]. An ABC-AIS hybrid approach to dynamic anomaly detection in AODV-based MANETs is found in [29]. Modified ABC algorithm based on fuzzy multi-objective technique for optimal power flow problem is presented in [30]. ABC optimization was used for multi-area economic dispatch in [31]. MMSE design of nonlinear volterra equalizers using ABC algorithm is introduced in [32]. Hybrid seeker optimization algorithm for global optimization is found in [33].

This paper proposes a hybrid LSSVM-ABC model. The performance of LSSVM is based on the selection of hyper parameters C (cost penalty), ϵ (insensitive-loss function) and γ (kernel parameter). ABC will be used to find the best parameter combination for LSSVM.

The paper is organized as follows: Section II presents the Least square support vector machine algorithm; Section III presents the Artificial Bee Colony algorithm; Section IV is devoted for the proposed model and implementation of ABC algorithms in stock prediction; In Section V the results are discussed. The main conclusions of the work are presented in Section VI.

## II. LEAST SQUARE SUPPORT VECTOR MACHINE

Least squares support vector machines (LSSVM) are least squares versions of support vector machines (SVM), which are a set of related supervised learning methods that analyze data and recognize patterns, and which are used for classification and regression analysis. In this version one finds the solution by solving a set of linear equations instead of a convex quadratic programming (QP) problem for classical SVMs. LSSVM classifiers, were proposed by Suykens and Vandewalle [34].

Let X is $n \times p$ input data matrix and $y$ is $n \times 1$ output vector. Given the $\{x_i, y_i\}_{i=1}^n$ training data set, where $x_i \in R^p$ and $y_i \in R$, the LSSVM goal is to construct the function $f(x) = y$, which represents the dependence of the output $y_i$ on the input $x_i$. This function is formulated as

$$f(x) = W^T \varphi(x) + b \qquad (1)$$

Where $W$ and $\varphi(x): R^p \to R^n$ are $n \times 1$ column vectors, and $b \in R$. LSSVM algorithm [8] computes the function (1) from a similar minimization problem found in the SVM method [6]. However the main difference is that LSSVM involves equality constraints instead of inequalities, and it is based on a least square cost function. Furthermore, the LSSVM method solves a linear problem while conventional SVM solves a quadratic one. The optimization problem and the equality constraints of LSSVM are defined as follows:

$$\min_{w,e,b} j(w,e,b) = \frac{1}{2} w^T w + C \frac{1}{2} e^T \qquad (2)$$

$$y_i = w^T \varphi(x_i) + b + e_i \quad (3)$$

Where e is the $n \times 1$ error vector, 1 is a $n \times 1$ vector with all entries 1, and $C \in R^+$ is the tradeoff parameter between the solution size and training errors. From (2) a Lagranian is formed, and differentiating with respect to $w, b, e, a$ ($a$ is Largrangian multipliers), we obtain

$$\begin{bmatrix} I & 0 & 0 & -Z^T \\ 0 & 0 & 0 & -1^T \\ 0 & 0 & CI & -I \\ Z & 1 & I & 0 \end{bmatrix} \begin{bmatrix} W \\ b \\ e \\ a \end{bmatrix} = \begin{bmatrix} 0 \\ 0 \\ 0 \\ y \end{bmatrix} \quad (4)$$

Where $I$ represents the identity matrix and

$$Z = [\varphi(x1), \varphi(x2), ..., \varphi(x_n)]^T \quad (5)$$

From rows one and three in (3) $w = Z^T a$ and $Ce = a$

Then, by defining the kernel matrix $K = ZZ^T$, and the parameter $\lambda = C^{-1}$, the conditions for optimality lead to the following overall solution

$$\begin{bmatrix} 0 & 1^T \\ 1 & K + \lambda I \end{bmatrix} \begin{bmatrix} b \\ a \end{bmatrix} = \begin{bmatrix} 0 \\ y \end{bmatrix} \quad (6)$$

**Kernel function K types are as follows:**

- Linear kernel $K(x, x_i) = x_i^T x$ (7)

- Polynomial kernel of degree d:
$K(x, x_i) = (1 + x_i^T x / c)^d$ (8)

- Radial basis function RBF kernel :

$$K(x, x_i) = \exp(-\|x - x_i\|^2 / \sigma^2) \quad (9)$$

- MLP kernel :

$$K(x, x_i) = \tanh(k x_i^T x + \theta) \quad (10)$$

### III. ARTIFICIAL BEE COLONY ALGORITHM

Artificial Bee Colony (ABC) algorithm was proposed by Karaboga in 2005 for real parameter optimization [35]. ABC is considered recent swarm intelligent technique. It is inspired by the intelligent behavior of honey bees. The colony of artificial bees consists of three groups of bees: employed, onlooker and scout bees. Half of the colony composed of employed bees and the rest consist of the onlooker bees. The number of food sources/nectar sources is equal with the employed bees, which means one nectar source is responsible for one employed bee. The aim of the whole colony is to maximize the nectar amount. The duty of employed bees is to search for food sources (solutions). Later, the nectars' amount (solutions' qualities/fitness value) is calculated. Then, the information obtained is shared with the onlooker bees which are waiting in the hive (dance area). The onlooker bees decide to exploit a nectar source depending on the information shared by the employed bees. The onlooker bees also determine the source to be abandoned and allocate its employed bee as scout bees. For the scout bees, their task is to find the new valuable food sources. They search the space near the hive randomly [36]. In ABC algorithm, suppose the solution space of the problem is D-dimensional, where D is the number of parameters to be optimized.

The fitness value of the randomly chosen site is formulated as follows:

$$\text{fit}_i = \frac{1}{(1 + \text{obj.fun}_i)} \quad (11)$$

The size of employed bees and onlooker bees are both SN, which is equal to the number of food sources. There is only one employed bee for each food source whose first position is randomly generated.. In each iteration of ABC algorithm, each employed bee determines a new neighboring food source of its currently associated food source and computes the nectar amount of this new food source by

$$v_{ij} = x_{ij} + \theta(x_{ij} - x_{kj}) \quad (12)$$

Where;

$$i = 1,2,\ldots,SN$$
$$j = 1,2,\ldots,D$$
$$\theta = \text{random number in range } [0,1]$$

If the new food source is better than that of previous one, then this employed bee moves to new food source, otherwise it continues with the old one.

After all employed bees complete the search process; they share the information about their food sources with onlooker bees. An onlooker bee evaluates the nectar information taken from all employed bees and chooses a food source with a probability related to its nectar amount by Equation:

$$p_i = \frac{\text{fit}_i}{\sum_{j=1}^{SN} \text{fit}_j} \quad (13)$$

Where $\text{fit}_i$ is the fitness value of the solution i which is proportional to the nectar amount of the food source in the position i and SN is the number of food sources which is equal to the number of employed bees.

Later, the onlooker bee searches a new solution in the selected food source site, the same way as exploited by employed bees. After all the employed bees exploit a new solution and the onlooker bees are allocated a food source, if a source is found that the fitness hasn't been improved for a predetermined number of cycles (limit parameter), it is abandoned, and the employed bee associated with that source becomes a scout bee. In that position, scout generates randomly a new solution by Equation:

$$x_i^j = x_{min}^j + r(x_{max}^j - x_{min}^j) \quad (14)$$

Where;

r is random number in range [0 , 1].

$x_{min}^j$ , $x_{max}^j$ are the lower and upper borders in the jth dimension of the problem space.

The ABC algorithm is shown in Fig. 1.

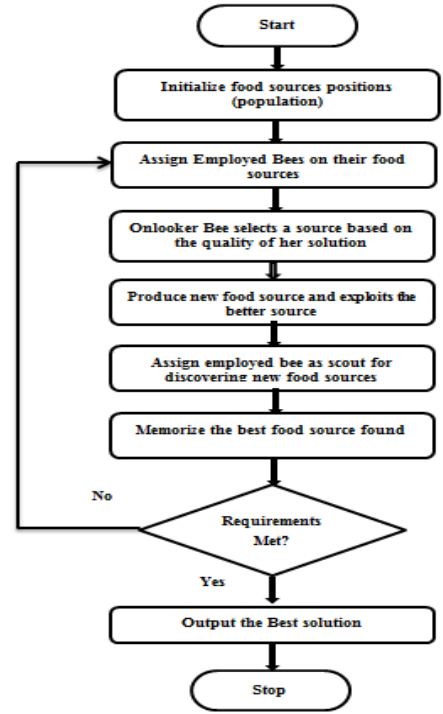

Fig.1 the ABC algorithm.

### IV. THE PROPOSED MODEL

The proposed model is based on the study of stock historical data (High, Low, Open, Close, and Volume). Then technical indicators are calculated from these historical data to be used as inputs to proposed model. After that LSSVM is optimized by ABC algorithm to be used in the prediction of daily stock prices. LSSVM optimized by PSO algorithm, standard LSSVM, and ANN trained with Scaled Conjugate gradient (SCG) algorithm, which is one of the best back-propagation derivatives, are used as benchmarks for comparison with proposed model. The proposed model architecture contains six inputs vectors represent the historical data and five derived technical indicators from raw datasets, and one output represents next price.

Proposed model phases are summarized in Fig 2.

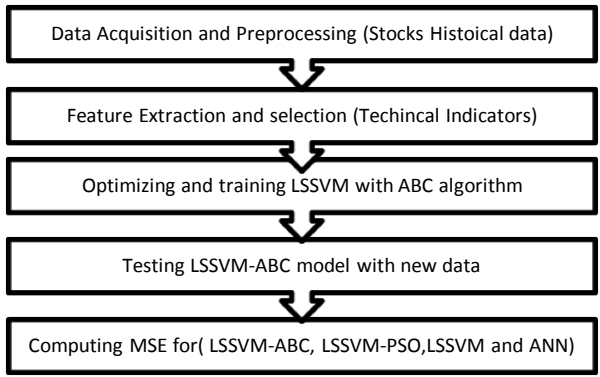

Fig. 2 the proposed model phases.

The technical indicators, which are calculated from the raw datasets, are as follows:

- **Relative Strength Index (RSI)**: A technical momentum indicator that compares the magnitude of recent gains to recent losses in an attempt to determine overbought and oversold conditions of an asset. The formula for computing the Relative Strength Index is as follows.

  RSI = 100- [100 / (1+RS)]     (15)

  Where RS = Avg. of x days' up closes divided by average of x days' down closes.

- **Money Flow Index (MFI)**: This one measures the strength of money in and out of a security. The formula for MFI is as follows.

  Money Flow (MF) = TP * V     (16)

  Where, TP is typical price, and V is money volume

  Money Ratio (MR) is calculated as:

  MR = (Positive MF / Negative MF)   (17)

  MFI = 100 – (100/ (1+MR)).     (18)

- **Exponential Moving Average (EMA)**: This indicator returns the exponential moving average of a field over a given period of time. EMA formula is as follows.

  EMA = [α *T Close] + [1-α* Y EMA].  (19)

  Where T is Today's close and Y is Yesterday's close.

- **Stochastic Oscillator (SO)**: The stochastic oscillator defined as a measure of the difference between the current closing price of a security and its lowest low price, relative to its highest high price for a given period of time. The formula for this computation is as follows.

  %K = [(CP – LP) / (HP –LP)] * 100   (20)

  Where, CP is Close price, LP is Lowest price, HP is Highest Price, and LP is Lowest Price.

- **Moving Average Convergence/Divergence (MACD)**: This function calculates difference between a short and a long term moving average for a field. The formulas for calculating MACD and its signal as follows.

  MACD= [0.075*E] - [0.15*E]     (21)

  Where, E is EMA (CP)

  Signal Line = 0.2*EMA of MACD  (22)

## V. RESULTS AND DISCUSSIONS

The proposed and compared models were trained and tested with datasets for twenty companies cover all sectors in S&P 500 stock market. Datasets period are from Jan 2009 to Jan 2012. All datasets are available in [37]. Datasets are divided into training part (70%) and testing part (30%).

LSSVM-ABC algorithm parameters are shown in table (1).

*Table (1) LSSVM-ABC algorithm parameters.*

| No. of bees | epochs | LSSVM kernel |
|---|---|---|
| 20 bees | 50 | RBF kernel |

LSSVM-PSO parameters are shown in table (2).

*Table (2) LSSVM-PSO algorithm parameters.*

| No. of Particles | epochs | LSSVM kernel |
|---|---|---|
| 20 particles | 50 | RBF kernel |

ANN parameters are found in table (3).

*Table (3) ANN model parameters.*

| Training algorithm | Input layer | Hidden layer | epochs | Output layer |
|---|---|---|---|---|
| SCG | 6 nodes | 11 nodes | 1000 | 1 node |

Fig. 3 outlines the ANN structure used in this paper.

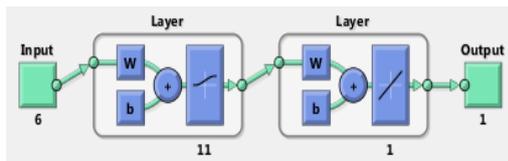

Fig. 3 ANN model structure

Fig. 4 to Fig. 23 outline the application of Proposed LSSVM-ABC model compared with LSSVM-PSO, LSSVM and ANN-BP algorithms at different datasets representing different stock market sectors.

In Fig. 4, Fig. 7, Fig. 8 and Fig. 9 which represent results of four companies in information technology sector (Adobe, Oracle, CISCO, and HP) results show that proposed LSSVM-ABC, and LSSVM-PSO models are achieving lowest error value with little advance to proposed model.

In Fig. 5, and Fig. 6 which represent results of two other companies in information technology sector (Amazon, and Apple), results show that ANN is fallen in overfitting problem, since the datasets have fluctuations. LSSVM-ABC algorithm is the best one with lowest error value and could easily overcome local minima and overfitting problems.

Fig. 10, and Fig. 11 which represent results of financial sector companies (American Express, and Bank of New york), we can remark that the predicted curve using the proposed LSSVM-ABC and LSSVM-PSO are most close to the real curves which achieve best accuracy, followed by LSSVM, while ANN-BP is the worst one.

Fig. 12 represents result of Honeywell company which represents industrials stock sector. Proposed model could easily overcome local minimum problem which ANN suffers from.

Fig. 13 and Fig.14 outline the application of the proposed algorithm to hospera and life technologies companies in health stock sector. From figures one can remark the enhancement in the error rate achieved by the proposed model.

Fig. 12, Fig.16, Fig.17, Fig.18, and Fig.19 outline the results of testing proposed model for (Coca-Cola, Exxon-mobile, AT&T, FMC, and duke energy) companies which represent different stock sectors. LSSVM-ABC and LSSVM-PSO can converge to global minima with lowest error value compared to LSSVM and ANN (especially in fluctuation cases).

Fig. 20, Fig.21, Fig.22, and Fig.23 represent results for (Ford, FEDX, MMM, and PPL) companies in different stock sectors. The results of proposed model has little advance compared with LSSVM and ANN since the data sets is normal and have no fluctuations, but LSSVM-ABC has fast convergence speed compared to other models.

Table 4 outlines Mean Square Error (MSE) performance function for proposed and compared models. It can be remarked that the LSSVM-ABC, and LSSVM-PSO techniques always gives an advance over LSSVM and ANN trained with SCG algorithms in all performance functions and in all trends and sectors.

Fig. 24 outlines comparison between LSSVM-ABC, LSSVM-PSO, LSSVM and ANN algorithms according to MSE function.

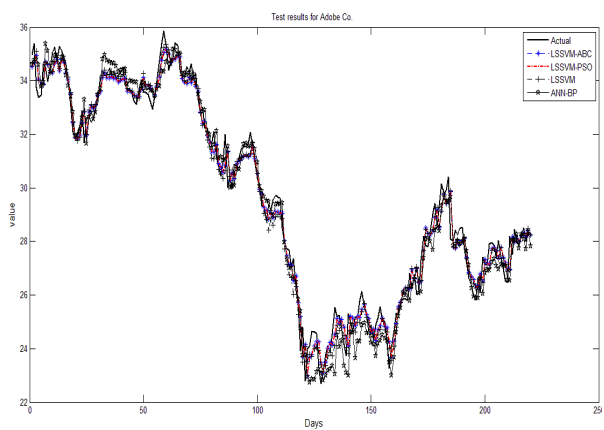

Fig. 4 Results for Adobe Company.

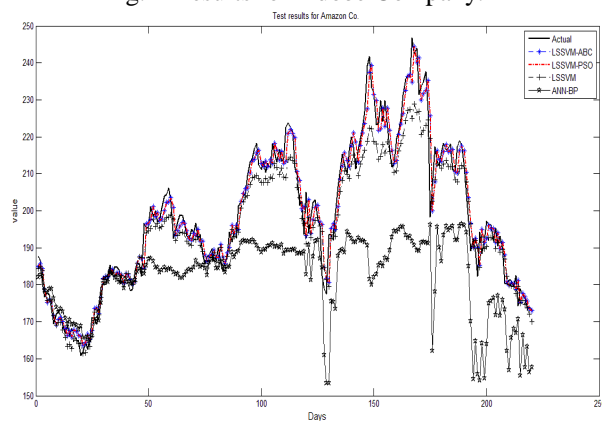

Fig. 5 Results for Amazon Company.

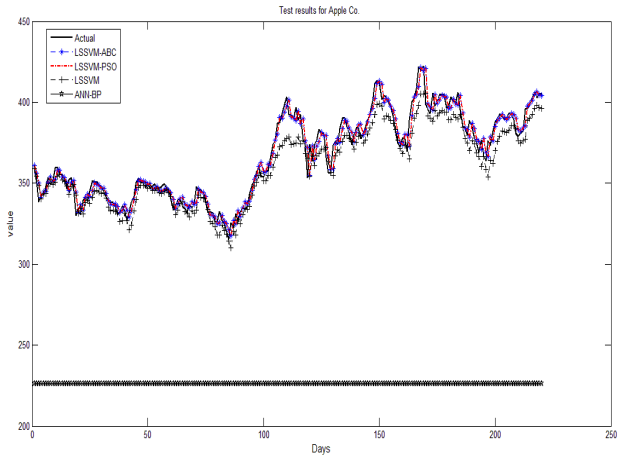

Fig. 6 Results for Apple Company.

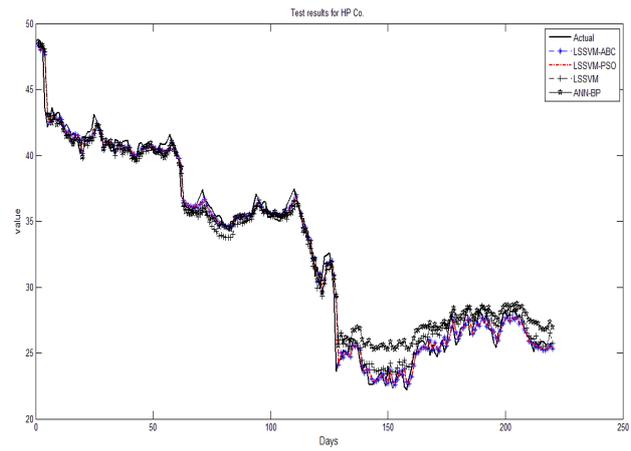

Fig. 9 Results for HP company.

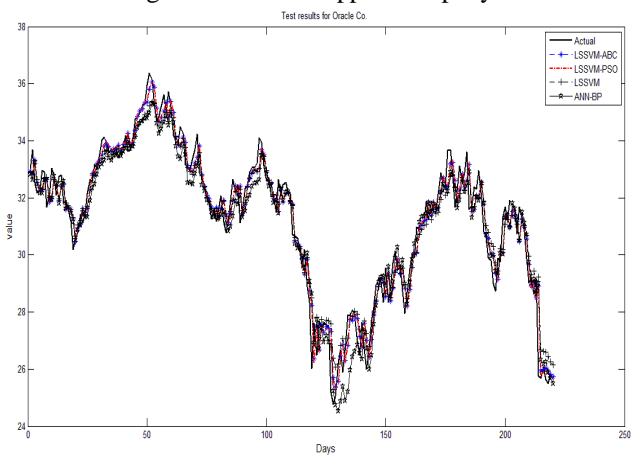

Fig. 7 Results for Oracle Co.

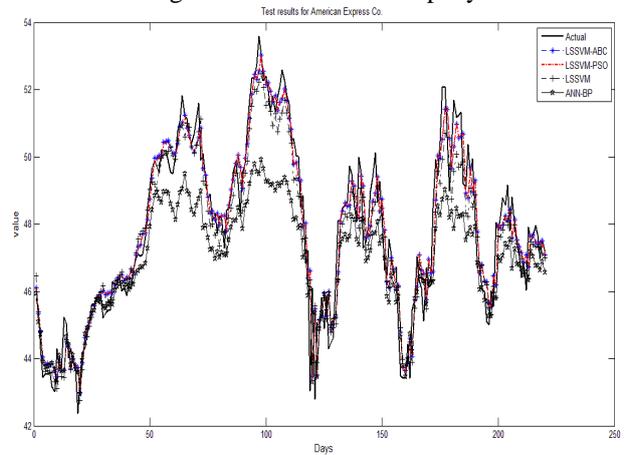

Fig. 10 Results for American Express Company

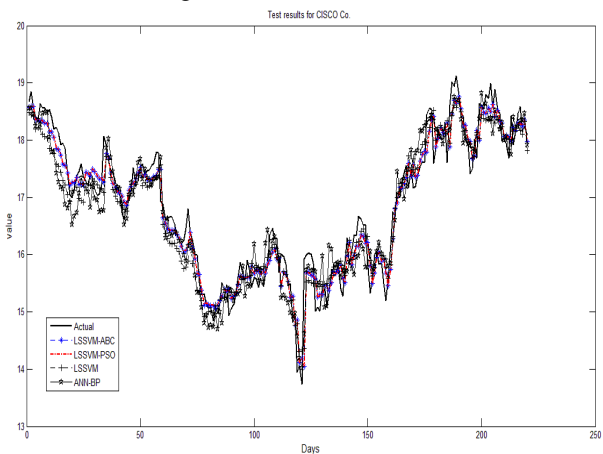

Fig. 8 Results for CISCO Co.

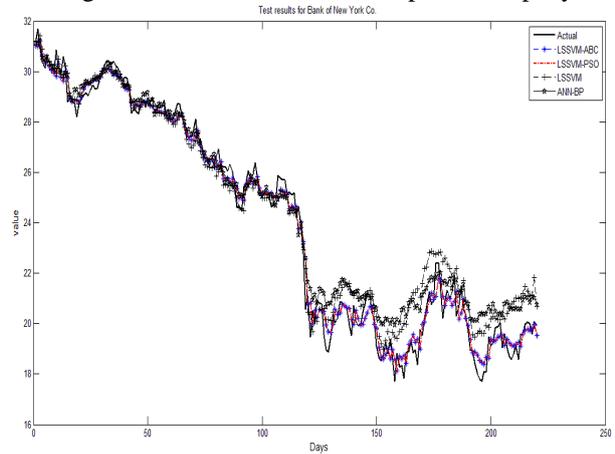

Fig. 11 Results Bank of New York company.

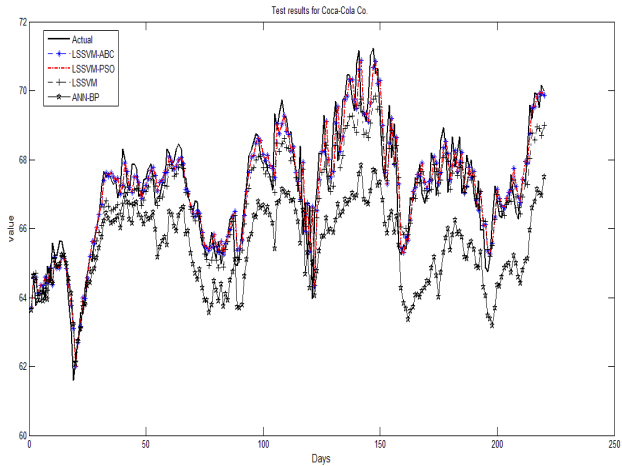

Fig. 12 Results for Coca-Cola company.

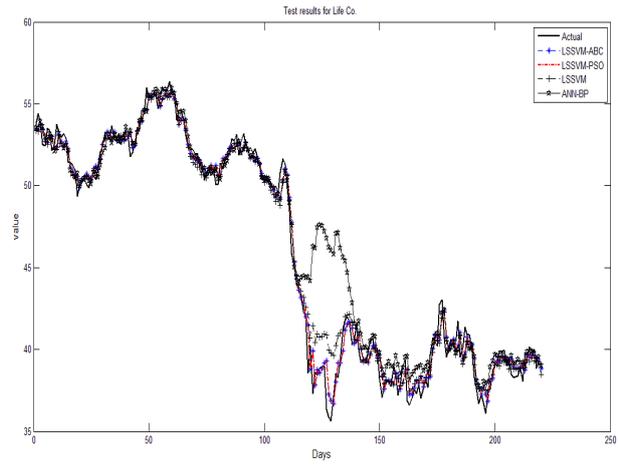

.Fig. 15 Results for Life Tech. Company

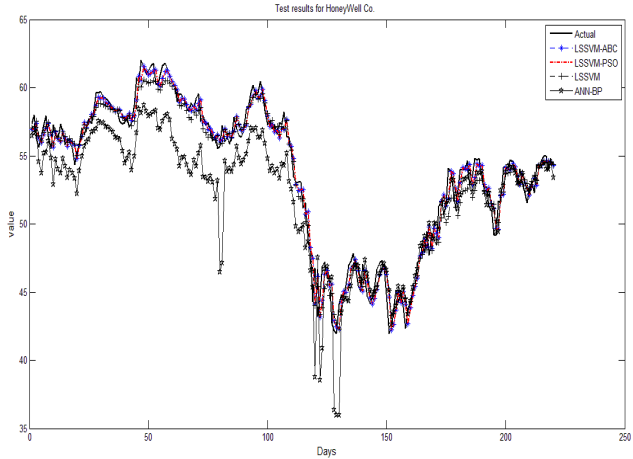

Fig. 13 Results for Honey Well company.

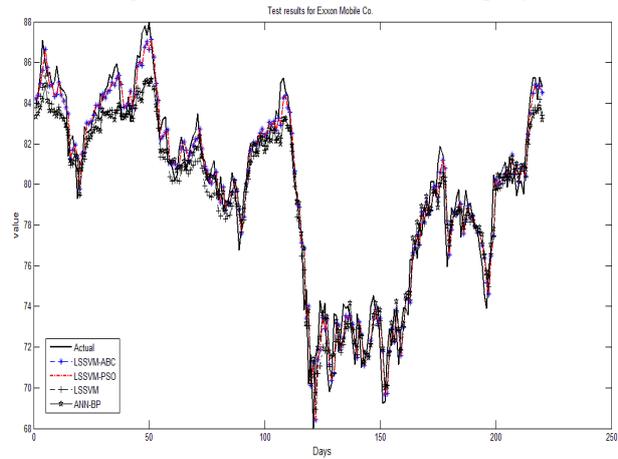

Fig. 16 Results for Exxon-Mobile. Company**.**

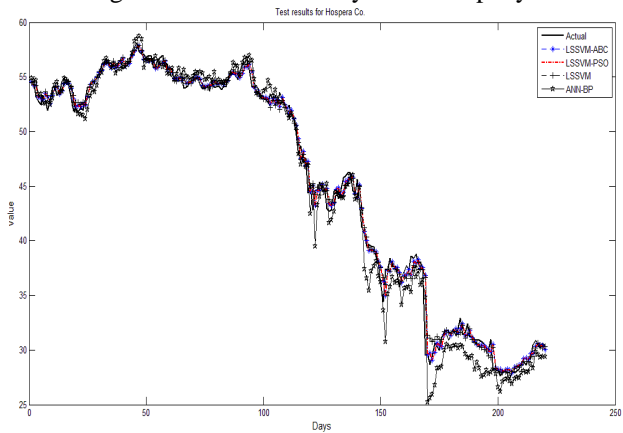

Fig. 14 Results for Hospera Co.

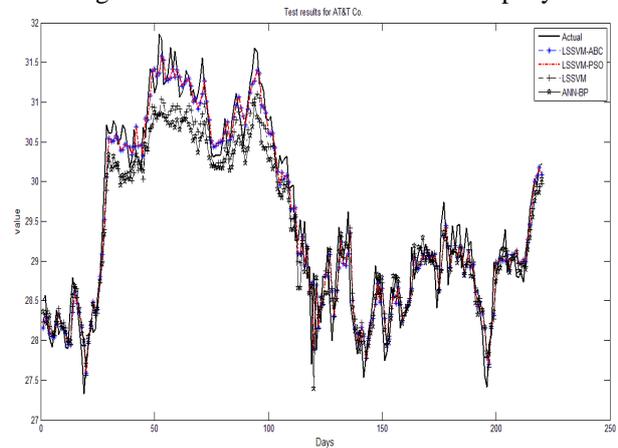

Fig. 17 Results for AT&T. Company

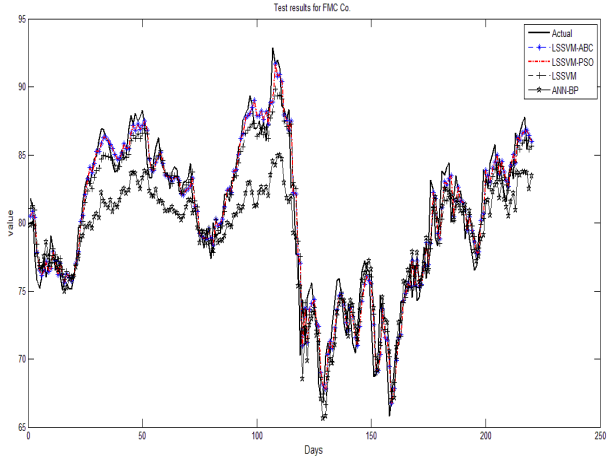

Fig. 18 Results for FMC. Company**.**

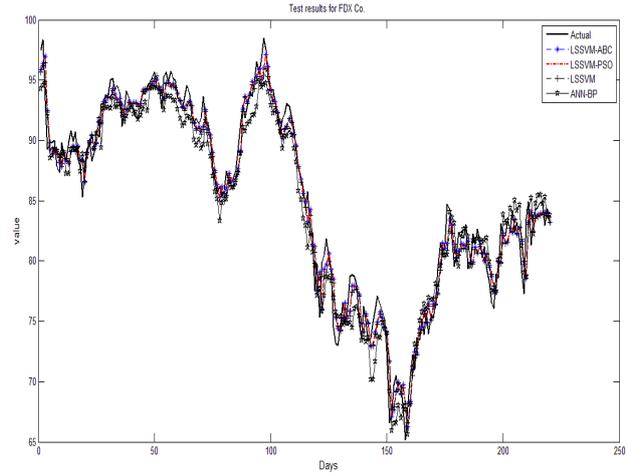

Fig. 21 Results for FEDX. Company**.**

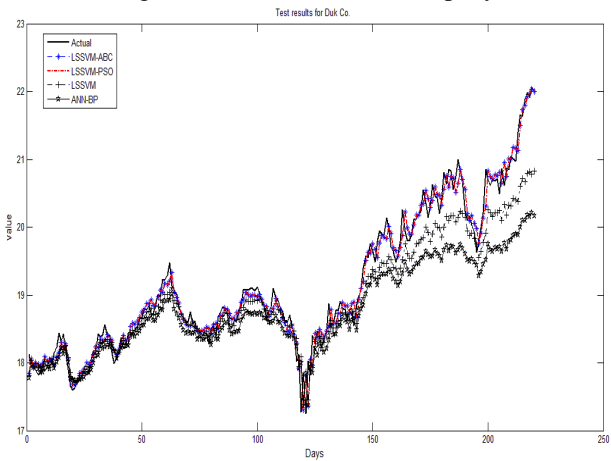

Fig. 19 Results for Duke. Company**.**

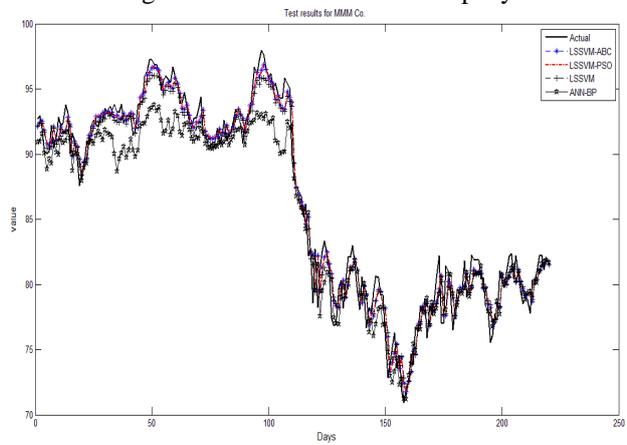

Fig. 22 Results for MMM. Company.

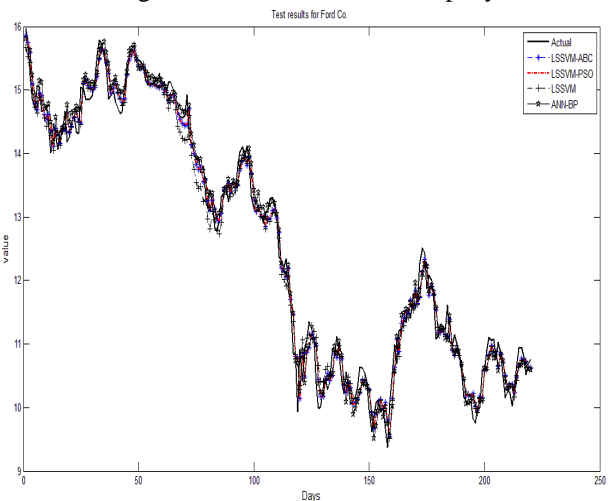

Fig. 20 Results for Ford. Company

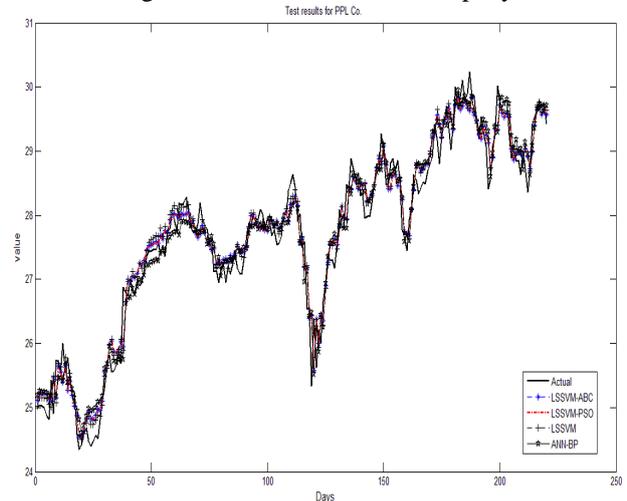

Fig. 23 Results for PPL. Company

*Table (4) Mean Square Error (MSE) for proposed algorithm.*

| Algorithm / Company | LSSVM-ABC | LSSVM-PSO | LSSVM | ANN |
|---|---|---|---|---|
| Adobe | 0.5310 | 0.5317 | 0.5529 | 0.6979 |
| Amazon | 4.5286 | 4.5315 | 6.1981 | 22.015 |
| Apple | 5.5915 | 5.5924 | 9.4863 | 143.01 |
| Oracle | 0.6320 | 0.6314 | 0.6807 | 0.7598 |
| CISCO | 0.3115 | 0.3111 | 0.3623 | 0.3929 |
| HP | 0.7727 | 0.7725 | 0.9752 | 1.3146 |
| Am.Expres | 0.7904 | 0.7905 | 0.8761 | 1.5116 |
| NY Bank | 0.4841 | 0.4839 | 0.9394 | 0.9646 |
| Coca-Cola | 0.6809 | 0.6823 | 0.8096 | 2.0560 |
| H. Well | 0.9563 | 0.9574 | 1.3371 | 2.1853 |
| Hospera | 0.8695 | 0.8694 | 0.8936 | 1.5162 |
| Life Tech. | 0.7718 | 0.7713 | 1.0195 | 2.4162 |
| Exx.Mob. | 1.0997 | 1.1000 | 1.3016 | 1.3080 |
| AT & T | 0.2916 | 0.2911 | 0.3684 | 0.4241 |
| FMC | 1.5881 | 1.5881 | 1.7529 | 3.0843 |
| Duke | 0.1710 | 0.1735 | 0.3709 | 0.5897 |
| FORD | 0.2473 | 0.2472 | 0.2660 | 0.2628 |
| FEDX | 1.4948 | 1.4948 | 1.5285 | 1.9454 |
| MMM | 0.7718 | 0.7713 | 1.0195 | 2.4162 |
| PPL | 0.2920 | 0.2919 | 0.2907 | 0.3073 |

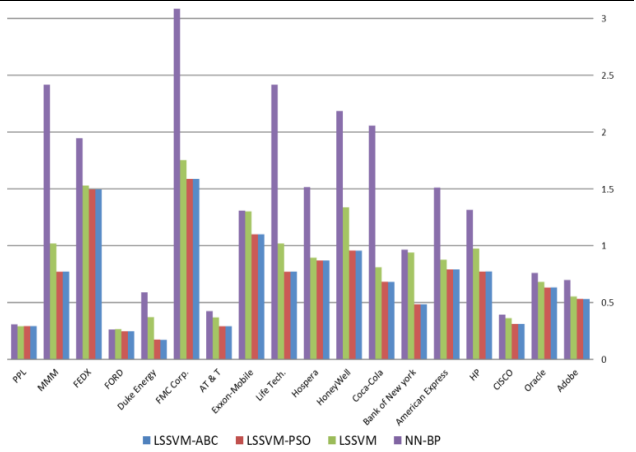

Fig. 24 MSE of proposed algorithm.

Convergence speed of the LSSVM-ABC, LSSVM-PSO, and ANN models are shown in Fig. 25, Fig. 26 and Fig. 27 respectively. From figures one can notice that the proposed model has fast convergence to global minimum after less than twenty iteration with lowest error value followed by LSSVM-PSO model, while ANN has worst convergence speed and has highest error value even after training for long period (one thousand iterations).

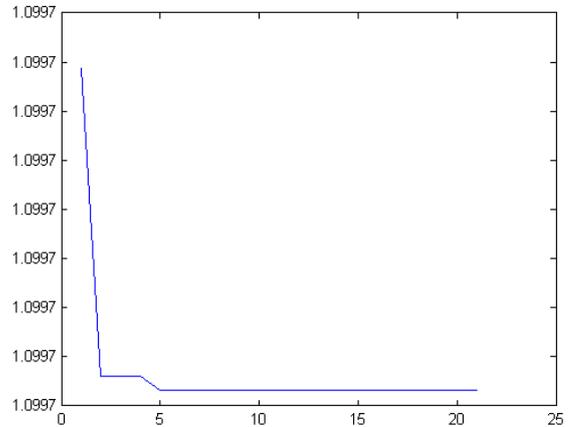

Fig. 25 Convergence curve of LSSVM-ABC.

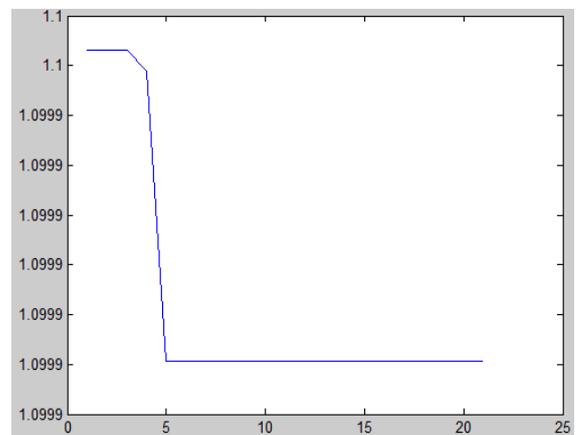

Fig. 26 Convergence curve of LSSVM-PSO.

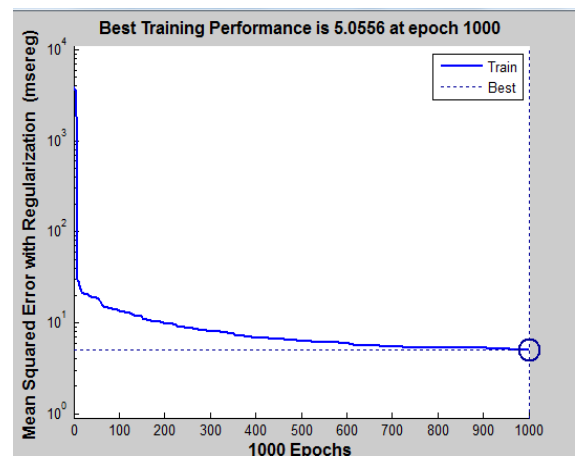

Fig. 27 Convergence curve of ANN model

## VI. CONCLUSION

Artificial Bee Colony algorithm (ABC), a population-based iterative global optimization algorithm is used to optimize LSSVM for stock price prediction. ABC algorithm is used in selection of LSSVM free parameters C (cost penalty), $\epsilon$ (insensitive-loss function) and $\gamma$ (kernel parameter). The proposed LSSVM-ABC model Convergence to a global minimum can be expected. Also proposed model overcome the over-fitting problem which found in ANN, especially in case of fluctuations in stock sector. LSSVM-ABC algorithm parameters can be tuned easily. Optimum found by the proposed model is better than LSSVM-PSO and LSSVM. Proposed model converges to global minimum faster than LSSVM-PSO model. LSSVM-ABC and LSSVM-PSO achieves the lowest error value followed by standard LSSVM, while ANN-BP algorithm is the worst one.